\documentclass{llncs}
\usepackage{graphicx}

\title{Clustering of scientific citations in Wikipedia}
\titlerunning{Clustering of scientific citations}
\author{Finn {\AA}rup Nielsen}
\institute{The Lundbeck Foundation Center for Integrated Molecular
  Brain Imaging \\
  at DTU Informatics and
  Neurobiology Research Unit, Copenhagen University Hospital
  Rigshospitalet 
  \\
  Lyngby and Copenhagen, Denmark
  \\
  \email{fn@imm.dtu.dk}\footnotetext{This document is licensed under
    Creative Commons CC-BY-SA,GNU General Public License or GNU Free
    Documentation License.}}

\begin{document}
\maketitle

\begin{abstract}
  The instances of templates in Wikipedia form an interesting data set
  of structured information. 
  Here I focus on the {\tt cite journal} template that is primarily
  used for citation to articles in scientific journals.
  These citations can be extracted and analyzed: Non-negative matrix
  factorization is performed on a (article $\times$ journal) matrix
  resulting in a soft clustering of Wikipedia articles and scientific
  journals, each cluster more or less representing a scientific topic.  
\end{abstract}

\section{Introduction}

The category system and the use of templates in Wikipedia provide 
interesting data sets of structured information.
A number of reports have come out that use the category graph in
automatic text processing, e.g., \cite{RuizCasadoM2005AutomaticAssignment,StrubeM2006WikiRelate,GabrilovichG2006Overcoming}.
DBpedia databases Wikipedia template information and associated
Internet services enable database-like queries
\cite{AuerS2008DBpedia}. 
I have previously reported results of relatively simple statistical
analysis about a single Wikipedia template --- the {\tt cite journal}
template --- counting the number of overall outbound scientific
citations and comparing it to the citation statistics {\em Journal
  Citation Reports} from the company {\em Thomson Scientific}
\cite{NielsenF2007Scientific}.    
Other researcher have considered more advanced statistical models in
the form of multivariate analysis
\cite{BuntineW2005Static,BellomiF2005Network}.
They build a matrix from intrawiki links and submit it to numerical
algorithms.   
Here I will take a similar approach but construct the matrix from data
associated with the scien\-tific citation template rather than wikilinks.
The present work will show an example on how to make multivariate
statistical analysis on the structured data in a Wiki, and in this
particular case provide an overview of how science is represented in
Wikipedia.

\section{Method: From XML via matrices to topic visualization}

A Perl script extracted the instances of the {\tt cite journal}
templates from bzip\-ped XML 
files of the English Wikipedia downloaded from the Internet server
{\tt download.wikipedia.org}. 
Another Perl script extracted the name of the journal from the {\tt
  journal} field in the template, and at the same time tried to match
the name to a `canonical' journal name.
For the matching a small XML file --- originally built for the
neuroinformatics Brede Database \cite{NielsenF2003Brede_abstract} ---
listed the canonical name and 
variations in the names for so far 255 different journals, e.g., the entry for
the journal with the canonical name {\em Proceedings of the Royal
  Society of London, Series B, Biological Sciences} listed 12 other
variations for the name including the PubMed abbreviation {\em Proc R
  Soc Lond B Biol Sci}. 
These 255 journals comprised a large part of the top cited journals
from 
Wikipedia, and thus the script normalizes very many citations to a
canonical name, but indeed far from all variations to lesser cited
journals are resolved. 
There are a number of other issues that prevents the databasing of the
citations to be particular exact: 
Special cases of journal naming make it hard to match all journal
names with a canonical journal name, e.g., 
{\em Mutation Research} are actually three (or four) different
journals, wrt. to ISSN. 
Cited `journals' may not be scientific journals,
but, e.g., newspapers.
Citations that occur multiple times in the same Wikipedia article to
the same item (by the {\tt <ref name="anchor"/>} construct) were only
counted once. 

A (article $\times$ journal) data matrix is built up where each column
corresponds either to a 
canonical journal name or the journal name as written in the citation of
the Wikipedia articles. 
Each row corresponds to a Wikipedia article. 
The $(i, j)$ elements in the matrix is set to the number of times the
$i$th article cites the $j$th journal. 
Most of the elements in the matrix are zero.

Clustering of the constructed matrix is performed with the
multiplicative update rules of the non-negative matrix factorization
(NMF) as put forward by Lee and Seung \cite{LeeDaniel2001Algorithms}.
The algorithm for the `Euclidean distance' runs with 50,000
iterations.
This particular multivariate analysis resembles several other methods
such as the one used by Buntine in his Wikipedia analysis
\cite{BuntineW2005Static} as well as Bellomi and coworkers' analysis
\cite{BellomiF2005Network}.  
NMF splits the data matrix $\bf X$(article $\times$ journal) into
three other matrices $\bf X = WH +U$.
Whereas $\bf U$ is just the residual matrix, the factorized matrices
$\bf W$ and $\bf H$ form the interesting matrices that may be expected
to represent specific 
scientific topics characterized by their citation patterns:
A specific column in $\bf W$ can be interpreted to contain the loadings
of articles on a specific `topic' that the cluster 
represents, and a specific row in $\bf H$ contains loadings for
journals on that topic.
The NMF results not in a hard clustering where the items are assigned
exclusively to one cluster, --- rather in a soft two-way clustering. 
Using the Kleinberg terminology \cite{KleinbergJon1997Authoritative},
the $\bf W$ matrix contains loadings for  Wikipedia `hub'
articles, whereas $\bf H$ contains `authoritative' journal articles.
One advantage of the Lee and Seung's `Euclidean distance' version of
the NMF algorithm is that no multiplications take place with the full
reconstructed data matrix, i.e., the product matrix $\bf WH$.
This is in contrast to `divergence' version, that in my implementation
is much slower and use more memory for these kinds of data sets.

The initialization of the NMF algorithm requires the specification of
the number of clusters, i.e., the number of columns in the $\bf W$
matrix and the number of rows in the $\bf H$ matrix. 
I make the NMF algorithm run with different number of
clusters: From one to twenty.
Each run will be independent of the other and they can be run in
parallel on a computer cluster.
Many results appear when running the NMF with different number of
clusters,
and a so-called `cluster bush' visualization can be used to get a
overview of the relationship between the different clusterings
\cite{NielsenF2005Mining}.
In this kind of plot each cluster is rendered as a circle and the
amount of overlap between two clusters is indicated with the thickness
of a line.

The NMF algorithm is run and the cluster bush visualization is made in
Matlab with functions from the Brede Toolbox
\cite{NielsenFinn2000Experiences}.

\section{Results}

\begin{figure}[bt]
  \centering
  \includegraphics[width=0.8\textwidth]{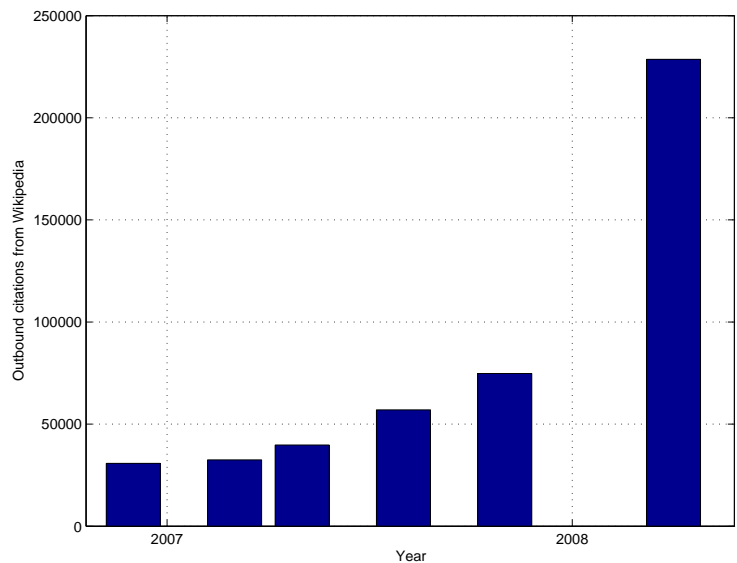}
  \caption{Number of outbound scientific citations as counted from the
    use of the {\tt cite journal} template for different dumps of the
    English Wikipedia. A sharp rise is seen from the 2007 dumps to the
  2008 dump due to citations added by a bot.}
  \label{fig:growth}
\end{figure}

Examing the full count of scientific citations from Wikipedia a marked
increase becomes apparent with a rise in the number of
citations from 2007 to the examined dump of March 2008, see
Figure~\ref{fig:growth}: 
From 74,776 citations in the October 2007 dump to 228,593 in the
March 2008 dump.

Whereas astronomy journals received comparably many citations from
Wiki\-pedia in the 2007 dumps, and journals such as {\em The Journal of
  Biological Chemistry} had relatively few citations when compared to
the {\em Journal Citation Reports}, this citation pattern is now very
much changed: 
Wikipedians have constructed the bot {\em ProteinBoxBot} that
automatically 
builds infoboxes and citations in Wikipedia articles. 
Thus a very large number of citations to protein/gene work has been added,
and with the March 2008 dump {\em The Journal of Biological Chemistry}
can be found as the most cited journal, see Table~\ref{tab:top}. 
Scientific articles cite also this journal the most, according to {\em
  Journal Citation Reports}.

\begin{table}[tb]
  \centering
  \begin{tabular}{rl}
    \hline\noalign{\smallskip}
    Citations & Journal name \\
    \hline\noalign{\smallskip}
    16739 & The Journal of Biological Chemistry \\
    12779 & PNAS \\
    8772 & Genome Research  \\
    7561 & Nature \\
    4007 & Nature Genetics     \\
    3928 & Genomics \\
    3689 & Science \\
    3511 & Gene \\
    3380 & Biochemical and Biophysical Research Communications \\
    3043 & Molecular and Cellular Biology \\
    2975 & Cell \\
    2261 & The EMBO Journal \\
    \hline\noalign{\smallskip}
  \end{tabular}
  \caption{Most cited journals from Wikipedia in the 12th March 2008
    dump.}
  \label{tab:top}
\end{table}

The conversion of the information in the templates to a matrix
representation results in matrices size $(23595 \times 18194)$ and
$(43073 \times 23096)$ for the October 2007 and March 2008 dump,
respectively. 
The densities of the constructed matrices are 0.01\%--0.02\% depending
on the dump version of Wikipedia.
The number of columns in the matrices would have been smaller and the
density higher if the matching of journal names was more complete.
The number of articles using the {\tt cite journal} template has
almost doubled in less than half a year between the two dumps.
This increase is likely due to the large number of articles added for
proteins/genes by {\em ProteinBoxBot}.
Many of these articles have no other text besides the text added by
the bot and the citations are not in-text citations. 

\begin{figure}[tb]
  \centering
  \includegraphics[width=\textwidth]{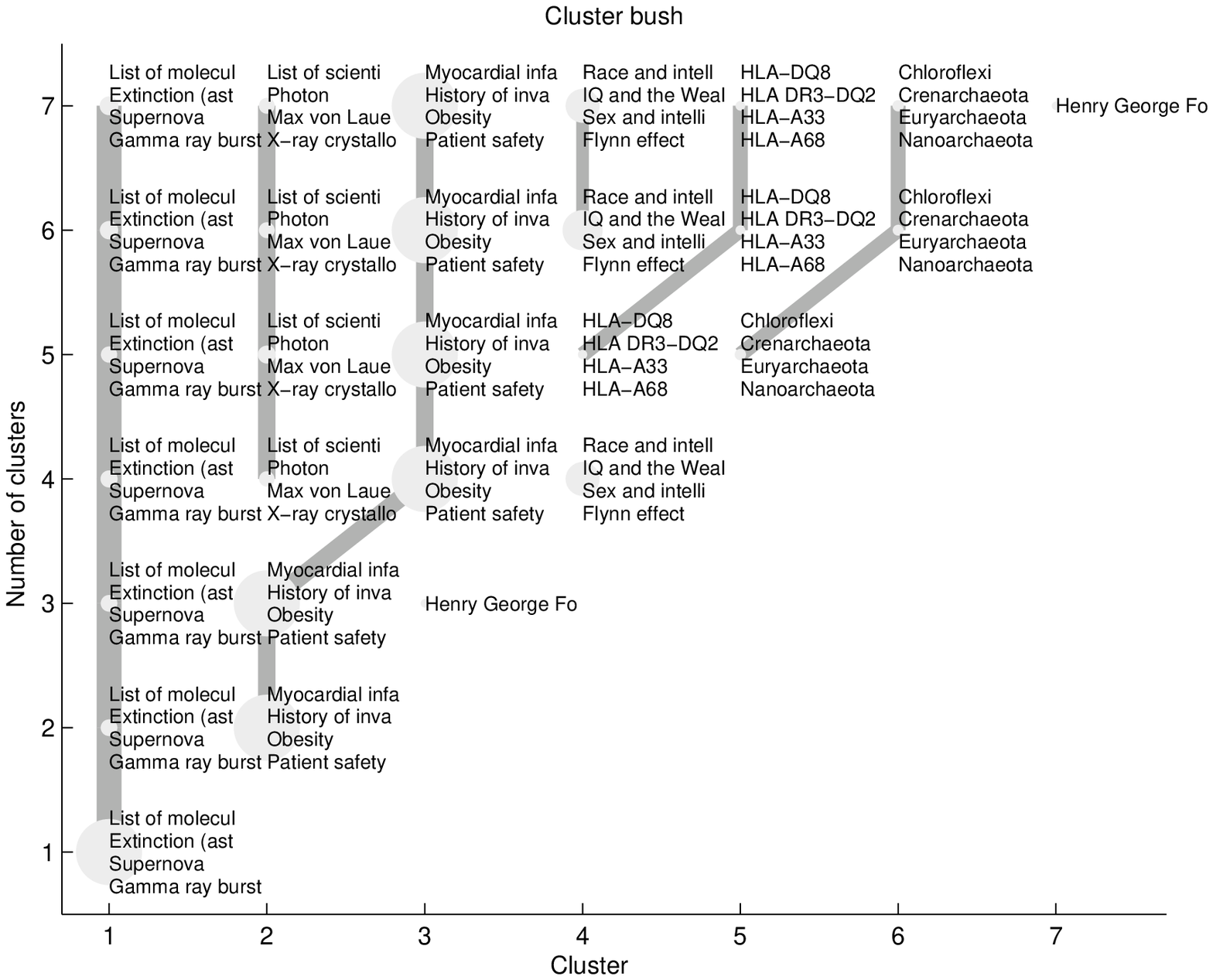}
  \caption{`Cluster bush' visualization of results for non-negative
    matrix factorization (NMF) of the scientific citations in the 18th
    October 2007 dump of the English Wikipedia.
    Each circle denotes a cluster.
    The lowest row displays the results of an NMF run with one
    cluster, second lowest row the results for NMF with two clusters,
    etc.
    The text on the nodes are the Wikipedia articles that are
    associated with high loadings in the factorized matrices of the NMF.
    The lines between the nodes indicate how much the clusters overlap.
  }
  \label{fig:clusterbush20071018}
\end{figure}

Figure~\ref{fig:clusterbush20071018} displays a cluster bush
visualization of the NMF results for the October 2007 dump for NMF,
and for clarity only the NMF runs with one to seven clusters are
shown: 
The bottom row displays the run with just one cluster, where the
Wikipedia articles {\em List of molecules in interstellar space} and
{\em Extinction (astronomy)} are the largest hubs. {\em The
  Astrophysical Journal} and {\em Astronomy \& Astrophysics} are the
largest authoritative journals for this astrophysical cluster. 
As the NMF model size increases, i.e., more clusters get added, this
topic continues to be a cluster of its own.
The new clusters that arise are related to medical sciences,
intelligence, human leukocyte antigen and bacteria.
For these runs of NMF the columns corresponding to the
cross-disciplinary journals {\em Nature}, {\em Science} and {\em
  PNAS} were excluded.

With the present algorithm a few of the clusters represent very
restricted topics, e.g., in one case the article {\em Henry George
  Fourcade} and the journal {\em The Photogrammetric Record}
constituted a single cluster.
Another cluster that is also dominated by single items has the article
about the group of genes {\em Solute carrier family} and the journal
{\em Pfl\"{u}gers Archiv European Journal of Physiology}.

Applying NMF on the March 2008 dump results in components that are
overwhelmingly affected by the large number of citations in the
protein/gene articles. 
A run of NMF with twenty clusters resulted in only three clusters that
did not exhibit an association with genes: 
One cluster centered around solar system astronomy with the journal {\em
  Icarus} as the primary authoritative journal and {\em Uranus} as the
top hub Wikipedia article, another cluster centered around {\em
  The Astrophysical Journal}, and the third as a medical clusters with
{\em New England Journal of Medicine} and {\em The Lancet} as top
authorities and {\em Myocardial infarction} as the Wikipedia hub
article. 
The rest of the seventeen clusters were all related to proteins and
genes or other closely related topics within biology and biochemistry.
Many of these clusters are mostly driven be a single journal, i.e., a
single element in each row of the $\bf H$ matrix are much larger than
the rest of the elements, whereas the $\bf W$ matrix shows a much more
equal loading over Wikipedia articles within each cluster, e.g., one
cluster interpretable as a `virology' cluster would have {\em The
  Journal of Virology} as the dominating authoritative journal. 

A few examples of items in a sample of clusters from an NMF run
with twenty clusters are shown in Table~\ref{tab:clusterexamples}.
These kinds of results may be written to an HTML page and put on the
web to serve as an online overview of how science is cited from
Wikipedia.

\begin{table}[tb]
  \centering
  \begin{tabular}{lll}
    \hline\noalign{\smallskip}
    Cluster & Wikipedia hub articles & Authoritative journals \\
    \hline\noalign{\smallskip}
    `Cancer' &  RBL2 & Oncogene \\
    & MYB  & Cancer Research \\
    & ERG (gene) & Int. J. Cancer \\
    & EPS8 & Gene \& Development \\
    \hline\noalign{\smallskip}
    `Immunology' & DNA vaccination & The Journal of Immunology \\
    & CCL21 & The Journal of Experimental Medicine \\
    & HLA-DQ8 & Tissue Antigens \\
    & HLA-DQA1 & Eur. J. Immunol. \\
    \hline\noalign{\smallskip}
    `Blood' & Acute myeloid leukemia & Blood \\
    & Serpin & British Journal of Haematology \\
    & CEBPE & The Journal of Clinical Investigation \\
    & CD34 & The Journal of Experimental Medicine \\
    \hline\noalign{\smallskip}
    `Virology' & Papillomavirus & The Journal of Virology \\
    & HHV Infected Cell \ldots & Virology \\
    & Poliovirus   & Journal of Molecular Biology \\
    & RELB         & AIDS Res. Hum. Retroviruses \\
    \hline\noalign{\smallskip}
  \end{tabular}
  \caption{The top Wikipedia hubs articles and authoritative journals
    with     respect to clusters from a non-negative matrix
    factorization with twenty clusters.}
  \label{tab:clusterexamples}
\end{table}


\end{document}